\documentclass[fleqn,usenatbib]{mnras}

\usepackage{newtxtext,newtxmath}

\usepackage[T1]{fontenc}

\DeclareRobustCommand{\VAN}[3]{#2}
\let\VANthebibliography\thebibliography
\def\thebibliography{\DeclareRobustCommand{\VAN}[3]{##3}\VANthebibliography}

\usepackage{graphicx}	
\usepackage{amsmath}	
\usepackage{lineno}
\usepackage{url}

\title[Strong scattering of PSR B1957+20]{Detection of strong scattering close to the eclipse region of PSR B1957+20}

\author[J. T. Bai et al.]
{J. T. Bai$^{1}$,
S. Dai$^{2}$\thanks{E-mail: Shi.Dai@westernsydney.edu.au (SD)},
Q. J. Zhi$^{1}$\thanks{E-mail: qjzhi@gznu.edu.cn (QJZ)},
W. A. Coles$^{3}$,
D. Li$^{4,5,6}$,
W. W. Zhu$^{4,5}$,
G. Hobbs$^{7}$,
G. J. Qiao$^{8}$,
N. Wang$^{9}$,
\newauthor J. P. Yuan$^{9}$,
M. D. Filipovic$^{2}$,
J. B. Wang$^{9}$,
Z. C. Pan$^{4,5}$,
L. H. Shang$^{1}$,
S. J. Dang$^{1}$,
S. Q. Wang$^{9}$,
C. C. Miao$^{4,5}$
\\
$^{1}$School of Physics and Electronic Science, Guizhou Normal University, Guiyang, 550001, People’s Republic of China\\
$^{2}$School of Science, Western Sydney University, Locked Bag 1797, Penrith South DC, NSW 2751, Australia\\
$^{3}$Department of Electrical and Computer Engineering, University of California, San Diego, La Jolla, CA 92093, USA\\
$^{4}$National Astronomical Observatories, Chinese Academy of Sciences, Beijing 100101, People’s Republic of China\\
$^{5}$University of Chinese Academy of Sciences, Beijing 100049, People’s Republic of China\\
$^{6}$NAOC-UKZN Computational Astrophysics Centre, University of KwaZulu-Natal, Durban 4000, South Africa\\
$^{7}$CSIRO Space and Astronomy, PO Box 76, Epping, NSW 1710, Australia\\
$^{8}$School of Physics, Peking University, Beijing 100871, People’s Republic of China.\\
$^{9}$Xinjiang Astronomical Observatory, 150, Science-1 Street, Urumqi, 830011 Xinjiang, People’s Republic of China
}

\date{Accepted XXX. Received YYY; in original form ZZZ}

\pubyear{2015}

\begin{document}
\label{firstpage}
\pagerange{\pageref{firstpage}--\pageref{lastpage}}
\maketitle

\begin{abstract}
We present the first measurement of pulse scattering close to the eclipse region of PSR B1957+20, which is in a compact binary system with a low-mass star. We measured pulse scattering time-scales up to 0.2\,ms close to the eclipse and showed that it scales with the dispersion measure (DM) excess roughly as $\tau\propto\Delta{\rm DM}^{2}$. Our observations provide the first evidence of strong scattering due to multi-path propagation effects in the eclipsing material. We show that Kolmogorov turbulence in the eclipsing material with an inner scale of $\sim100$\,m and an outer scale of the size of the eclipse region can naturally explain the observation. Our results show that the eclipsing material in such systems can be highly turbulent and suggest that scattering is one of the main eclipsing mechanisms at around 1.4\,GHz.
\end{abstract}




\section{Introduction}\label{sec1} 

Eclipses of radio emission from pulsars are commonly observed in compact binary systems consisting of a millisecond pulsar (MSP) and a low-mass star ($<1$\,$M\odot$). Such eclipses are generally believed to be caused by materials in the orbit interfering with the propagation of the radio emission and several eclipsing mechanisms have been proposed in the literature~\citep[see][for a review]{thompson94}. Thanks to extensive studies of a number of eclipsing pulsars, such as PSRs B1957+20~\citep{fruchter88b,fruchter90,Polzin20}, J1227-4853~\citep{roy15,Kudale20}, J1744+4937~\citep{lyne90},  J2051-0827~\citep{Stappers98,Polzin19}, J1544+4937~\citep{bhattacharyya13}, PSR J1810+1744~\citep{Polzin18}, our understanding of the eclipsing mechanism and properties of eclipsing materials have been greatly improved. Recently, \citet{kansabanik21} presented multi-frequency observations of PSR J1544+4937 and modelled its broadband radio spectrum. Their results suggested that synchrotron absorption by relativistic electrons is the dominating eclipsing mechanism at $\lesssim1$\,GHz.

As the first eclipsing pulsar discovered~\citep{fruchter88a}, PSR B1957+20 and its companion have been studied in great detail at multiple wavelengths from radio~\citep[e.g.,][]{fruchter90,Polzin20} to optical~\citep[e.g.,][]{fruchter88b,kulkarni88,paradijs88,kerkwijk11} and X-ray~\citep[e.g.,][]{sgk+03}. The detection of plasma lensing effect close to the eclipse makes PSR B1957+20 unique to probe not only eclipsing materials but also pulsar radiation mechanisms~\citep[e.g.,][]{Main17,Main18,Mahajan18,llm+19}. Early observations with the Arecibo telescope showed that the eclipse duration is frequency ($\nu$) dependent and roughly follows $\nu^{-0.41}$~\citep{fruchter90,rt91}. Before and after the eclipse, the pulsar signal is delayed by as much as 400\,$\mu$s, which can be explained by an excess of ionized medium with a column density of free electrons of $\sim10^{17}$\,cm$^{-2}$~\citep{fruchter90,rt91}. Recently, \citet{Polzin20} presented a study of PSR B1957+20 at low radio frequencies, and their measurements of eclipse duration and dispersion measure (DM) excess generally agree with previous results. 

Based on early observations, \citet{thompson94} examined a range of eclipsing mechanisms for PSR~B1957+20. They concluded that a favoured model is cyclotron or synchrotron absorption by plasma embedded in the pulsar wind combined with pulse smearing at high frequency (i.e. 1.4\,GHz). However, the cyclotron absorption mechanism is challenged by tight upper limits on the strength of magnetic fields near the eclipse obtained through studies of plasma lensing~\citep{llm+19}. \citet{Polzin20} presented simultaneous measurements of pulsed and continuum flux densities of PSR B1957+20 at 150\,MHz and showed that the pulsar signal is removed from the line of sight throughout the main body of the eclipse, which supports absorption or nonlinear scattering eclipse mechanisms. While \citet{Polzin20} observed some evidence of pulse profile broadening at the edge of eclipse region, whether scattering is playing an important role is inconclusive. Previously, strong scattering has only been observed in PSR B1259$-$63 close to its eclipse and was suggested to originate from plasma density fluctuations in the disk of its companion Be star~\citep{jml+96}. \citet{ymc+18} observed depolarisation of radio emission from PSR~J1748$-$2446A at particular orbital phases and suggested that this depolarisation occurs because of rotation-measure fluctuations resulting from turbulence in the stellar wind.

While several eclipsing mechanisms, such as cyclotron/synchrotron absorption, pulse smearing, and scattering, are strong at low frequencies, it is challenging to distinguish them from each other at low frequencies. High signal-to-noise ratio (S/N) observations at high frequencies are therefore extremely valuable for us to get a wideband understanding of these mechanisms and compare them with theoretical predictions. Because of its steep spectrum~\citep[e.g.,][]{fruchter90}, previous observations of PSR~B1957+20 at high frequencies (e.g., 1.4\,GHz) have limited sensitivities to probe various eclipsing mechanisms. In this paper we present multiple observations of the eclipse of PSR B1957+20 centred at $\sim1.25$\,GHz using the Five-hundred-meter Aperture Spherical Telescope~\citep[FAST,~][]{nan11,Li18,Jiang19}. We present the first measurement of pulse scattering close to the eclipse and new measurements of the eclipsing duration and aim to shed new light on the eclipsing mechanism and properties of eclipsing materials. Details of observations and data processing are provided in Section~\ref{sec2}. In Section~\ref{sec3}, we present our measurements of the frequency dependence of the eclipse duration and pulse scattering. Section~\ref{sec4} gives discussions and conclusions.  

\section{OBSERVATIONS AND DATA PROCESSING}\label{sec2} 

Three observations of PSR~B1957+20 were taken using the central beam of the 19-beam receiver of FAST. Observations on 2019 August 8 (2.25\,hr) and 2020 January 22 (3.5\,hr) covered the whole eclipsing period while the observation on 2020 January 27 (2.4\,hr) only covered a fraction of the eclipse and the egress. Data were recorded with the Reconfigurable Open Architecture Computing Hardware generation 2 (ROACH2) \footnote{https://casper.berkeley.edu/} backend in the pulsar search-mode with 8-bit sampling and 4096 channels within a frequency range from 1050 to 1450\,MHz. Only the total intensity were recorded and therefore we do not have full polarisation information. A time resolution of either 98.304 or 49.152 $\mu$s were used for these observations. 

Search-mode data were folded with \textsc{DSPSR}~\citep{vanstraten11} using pulsar parameters obtained from the Australian Telescope National Facility Pulsar Catalog ({\sc psrcat}~V1.64\footnote{https://www.atnf.csiro.au/research/pulsar/psrcat/})~\citep{Manchester05}. Because of variations in orbital parameters often observed in eclipsing binary systems~\citep[e.g.][]{arzoumanian94,shaifullah16}, we observed significant pulse drifting in phase with these parameters, which reduces the S/N of detection. In order to maximize the S/N of folded profiles, we re-folded each epoch with a refined set of parameters. These refined parameters were determined, for each epoch, by measuring time of pulse arrivals (ToAs) away from the eclipse and refitting for the spin frequency and its derivative ($F0$ and $F1$), the orbital period ($P_{\rm B}$) and the time of ascending node ($T0$) using \textsc{TEMPO2}~\citep{HEM06}\footnote{We note that these parameters can be highly co-variant since our integration time is much shorter than the orbital period. However, this will not affect our purpose of minimising pulse smearing due to the variation of orbital parameters.}. 
Refolded observations have 1\,s sub-integration time, 256 pulse phase bins and 1024 frequency channels (corresponding to a frequency resolution of 488.3\,kHz). Strong radio-frequency interference (RFI) were identified and masked using the \textsc{paz} and \textsc{pazi} commands of \textsc{PSRCHIVE}\footnote{https://psrchive.sourceforge.net/}~\citep{HVM04}. Flux densities were measured using the \textsc{PSRFLUX} command of \textsc{PSRCHIVE} with a standard template \citep[see e.g.,][for details]{Kumamoto2021}. A series of standard template was generated as a function of the orbital phase (every 5\,minutes) in order to account for changes in the shape of pulse profile due to smearing and scattering. 

To measure the DM as a function of the orbital phase, we further averaged the folded pulse profile to have a time resolution of 30\,s and a frequency resolution of 50\,MHz. As we go deeper into the eclipse region, the DM excess becomes large. In order to minimize the DM smearing, for each 30\,s of data, we first used the \textsc{pdmp} command of \textsc{PSRCHIVE} to obtain a rough estimate of the DM and then re-folded the search-mode data with the updated DM. A standard template was generated using the re-folded pulse profile. ToAs were then measured for each channel and the DM was measured using \textsc{TEMPO2}. 

The duration of eclipse was measured following the method described in \citet{Polzin20} and \citet{broderick16}. Measured flux densities were normalised so that the mean flux density of out-of-eclipse region was uniform. The eclipse ingress and egress flux densities, $f$, as a function of the orbital phase, $\phi$, were fitted to a Fermi-Dirac type function, $f=[e^{\frac{\phi -p_{1}}{p_{2}}}+1]^{-1}$, where $p_1$ is the orbital phase where the flux is half the out-of-eclipse value and $p_2$ is the slope.
Flux densities were measured every 5\,s using pulse profiles with updated DMs.

To search for evidence of pulse scattering and measure the scattering time-scale, we assumed that the observed pulse profile close to the eclipse region is the convolution of a standard template with a one-sided exponential decay function~\citep[e.g.,][]{lewandowski13}, 
\begin{equation}
s(t) = {\rm exp}(\frac{-t}{\tau_{\rm s}})H(t)
\label{eq2}
\end{equation}
where $\tau_{\rm s}$ is the pulse scattering time-scale and $H(t)$ is equal to zero for $t<0$, and one otherwise. This assumes a thin scattering screen located in between the pulsar and the observer. The standard template was formed using a high S/N pulse profile away from the eclipse. We computed the convolution numerically and fitted for the pulse scattering time-scale using the \textsc{bilby} \footnote{https://lscsoft.docs.ligo.org/bilby/} package in python~\citep{ahl+19}. \textsc{bilby} is a user-friendly Bayesian inference library to perform parameter estimation. The pulse scattering time-scale was measured for each 60\,s window covering orbital phases from 0.22 to 0.29. We assumed a Gaussian-likelihood function with a uniform prior for each parameter. The uncertainty of $\tau_{\rm s}$ was estimated as the average of deviations from the median at 16 per cent and 84 per cent of its posterior distribution.

\begin{figure*}
\centering
\includegraphics[width=180mm]{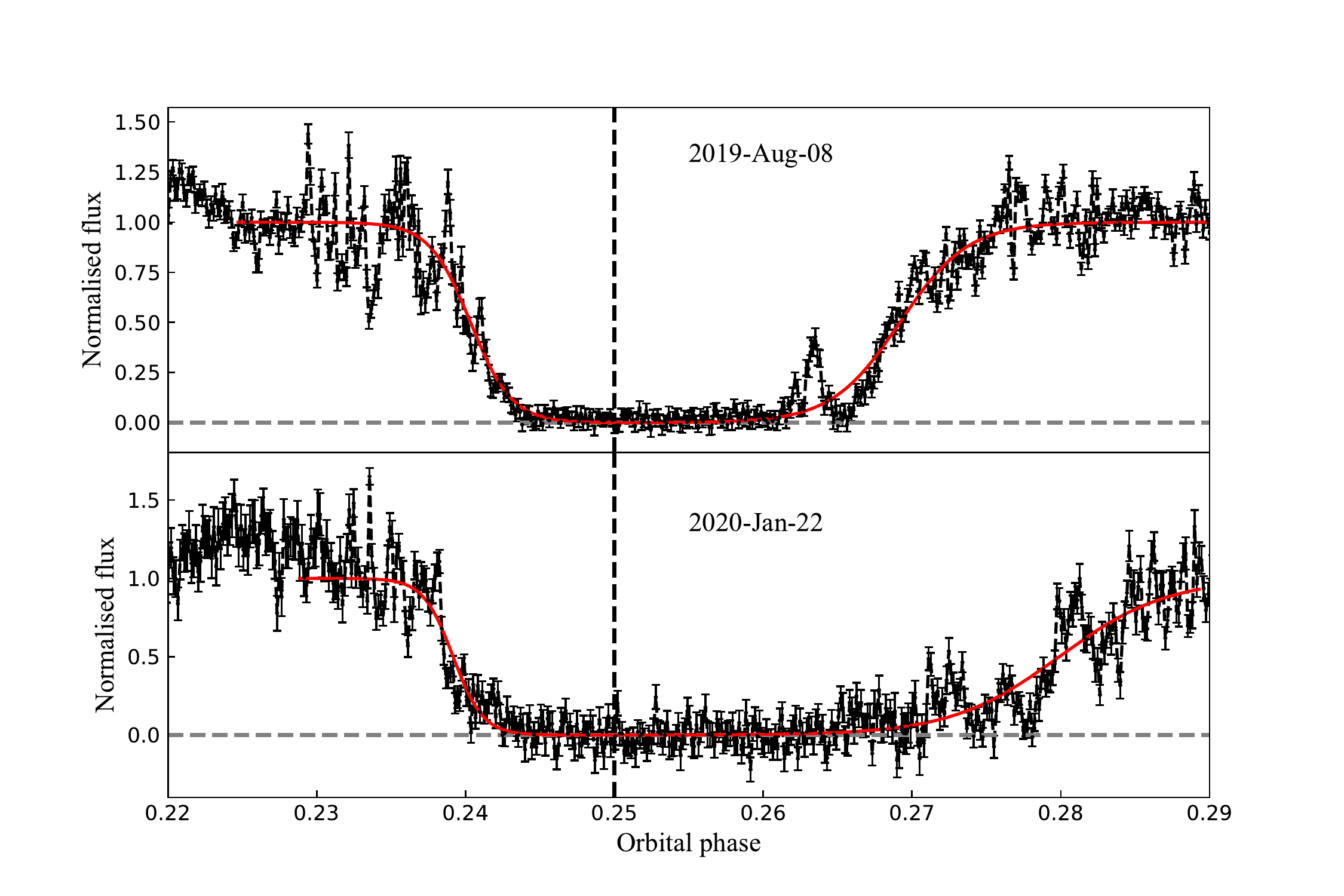}
\caption{Normalized flux density of pulse emission as a function of orbital phase. Red lines show the least-squares fitting of a Fermi–Dirac function to the ingresses and egresses of eclipse.
}
\label{ecl_fit}
\end{figure*}

\section{Results}\label{sec3} 

\subsection{Eclipse duration}\label{sec3.1}

Normalised flux densities measured on 2019 August 8 (top panel) and 2020 January 22 (bottom panel) are shown in Fig.~\ref{ecl_fit} as a function of the orbital phase. While the light curve during the ingress is similar for the two observations, it is significantly different during the egress, which has also been reported by previous observations~\citep{fruchter90,rt91,Polzin20}. We detected a sudden increase of flux density at orbital phases from $\sim0.261$ to $0.265$ on 2019 August 8, which is well within the eclipse region that previous observations suggested and has not been observed before. 

\begin{figure}
\centering
\includegraphics[width=90mm]{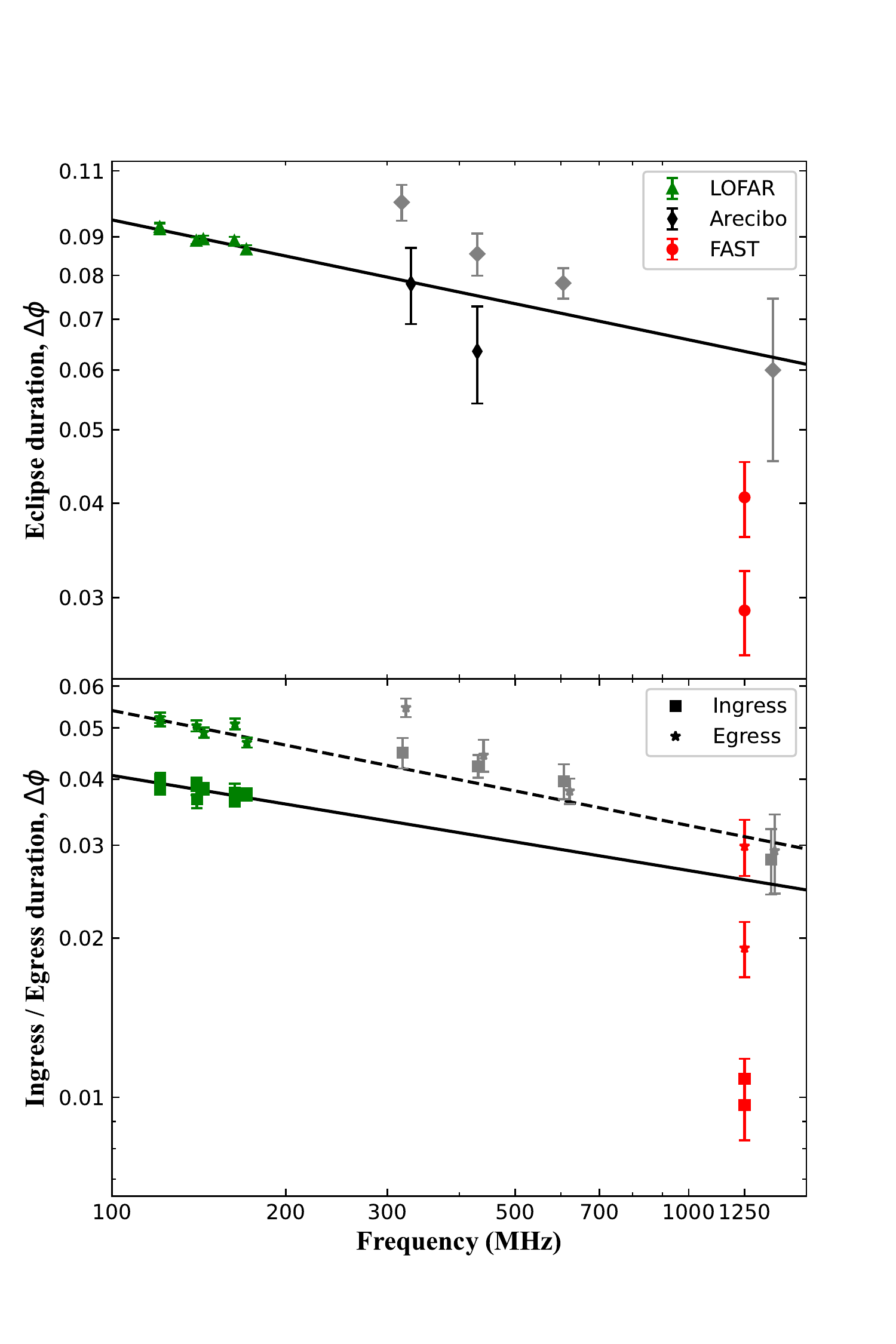}
\caption{Top: total eclipse duration as a function of the observing frequency. 
Green, black and gray points are published measurements from \citet{Polzin20} and \citet{rt91}. Red points are measurements using FAST observations on 2019 August 8 and 2020 January 22. Bottom: Ingress and egress duration as a function of the observing frequency. Colours are consistent with the top panel. In both panels, solid and dashed lines show the best-fit power-law model from \citet{Polzin20}. 
}
\label{ecl_time}
\end{figure}

We measured the duration of eclipse as described in Secion~\ref{sec2}. The duration of eclipse ingress and egress ($\Delta \phi _ {\rm in}$ and $\Delta \phi _ {\rm eg}$) are defined as the orbital phase difference between ingress/egress and inferior conjunction of the companion, $\phi = 0.25$. The total eclipse ($\Delta \phi _ {\rm eclipse}$) is defined as the orbital phase difference between ingress and egress measured at 1250\,MHz. Fig.~\ref{ecl_time} shows measured eclipse duration as a function of the observing frequency. For comparison, we also plotted recent low-frequency measurements from ~\citet{Polzin20} (green points) and previous Arecibo measurements~\citep[grey and black points][]{rt91,Polzin20}.

\begin{table*}
\centering
\caption{Measurements of the scattering time-scale ($\tau _{\rm s}$), the scattering index ($\alpha$) and the DM excess ($\Delta \rm DM$).}
{\begin{tabular}{cccccccc}
\hline
\hline
Date of obs.  &  Orbital phase  	&  \multicolumn{4}{c}{$\tau _{\rm s}$($\rm \mu s$)} 	& $\alpha $   		&  $\Delta \rm DM$  	        	\\
             &          &   1100 \rm MHz & 1200 \rm MHz & 1300 \rm MHz & 1400 \rm MHz       &       &   (\rm pc cm$^{-3}$)          \\
\hline	
2019-Aug-08 & & & & & & &\\
    &  0.2326 		&	19.5$^{+0.4} _{-0.4}$    & 14.5$^{+0.9} _{-0.8}$  &     &        &      	 &  0.035$\pm$0.001  	\\
    &  0.2344 		&	15.5$^{+0.3} _{-0.3}$    & 12.9$^{+0.9} _{-0.8}$  & 8$^{+2} _{-1}$   & 14$^{+2} _{-2}$ &  	&  0.035$\pm$0.001  	\\	
    &  0.2362    	&   40.5$^{+0.6} _{-0.5}$    & 33.9$^{+1.0} _{-0.8}$  & 31$^{+2} _{-2}$  & 30$^{+3} _{-3}$ &   -1.6$\pm$0.3    	&  0.046$\pm$0.002 		\\
    &  0.2381 		&	45.8$^{+0.5} _{-0.6}$    & 36$^{+1} _{-1}$  & 32$^{+1} _{-1}$  & 20$^{+2} _{-3}$ &	 -2.6$\pm$0.4  	&  0.057$\pm$0.003  	\\	
    &  0.2399    	&   84$^{+2} _{-2}$    & 90$^{+3} _{-3}$  & 68$^{+4} _{-3}$  & 68$^{+6} _{-7}$ &       	&  0.078$\pm$0.003		\\
    &  0.2417 		&	160$^{+7} _{-7}$   & 105$^{+6} _{-6}$ & 136$^{+8} _{-9}$ &	                   &  	&  0.13$\pm$0.01  	\\	
    &  0.2617    	&   138$^{+5} _{-5}$    & 111$^{+4} _{-4}$   & 94$^{+8} _{-9}$   &	                       &   -2.4$\pm$0.1       &  0.08$\pm$0.01 		\\
    &  0.2635 		&	153$^{+4} _{-3}$    & 103$^{+9} _{-9}$   & 104$^{+9} _{-11}$ &	                       &  	&  0.10$\pm$0.04  	\\	
    &  0.2653    	&   184$^{+14} _{-12}$  & 133$^{+15} _{-18}$ &                         &	             &        &  0.083$\pm$0.007 		\\
    &  0.2671 		&	77$^{+1} _{-2}$     & 54$^{+1} _{-2}$    & 72$^{+3} _{-3}$   & 31$^{+6} _{-6}$     &	  	&  0.053$\pm$0.006  	\\	
    &  0.2690    	&   29.4$^{+0.3} _{-0.4}$     & 23.2$^{+0.8} _{-0.9}$    & 21$^{+1} _{-1}$   &                           &   -2.3$\pm$0.4    	&  0.045$\pm$0.004 		\\
    &  0.2708 		&	21.9$^{+0.3} _{-0.4}$     & 18.6$^{+0.6} _{-0.6}$    & 15$^{+1} _{-1}$   &                           &	 -2.2$\pm$0.4  	&  0.046$\pm$0.006      \\	
    &  0.2726    	&   18.0$^{+0.3} _{-0.4}$     & 10.4$^{+0.7} _{-0.7}$    &     &    	&    		&  0.030$\pm$0.002 	\\
    &  0.2744 		&	9.0$^{+0.3} _{-0.3}$      & 7.9$^{+0.6} _{-0.6}$     &     &	    &	  		&  0.021$\pm$0.001  \\	
\hline
2020-Jan-22 & & & & & & &\\
    &  0.2344 		&	28.0$^{+0.9} _{-0.9}$    & 19$^{+1} _{-1}$  &     &	    &	 		&  0.041$\pm$0.002  \\	
    &  0.2362    	&   38$^{+1} _{-1}$    & 16$^{+1} _{-1}$  & 27$^{+2} _{-2}$   &                                &      	&  0.043$\pm$0.002   	\\
    &  0.2381 		&	24.3$^{+0.9} _{-1.0}$    & 24.5$^{+1.0} _{-0.9}$  & 14$^{+3} _{-3}$   &	                             & 	&  0.051$\pm$0.006 	    \\	
    &  0.2399    	&   167$^{+16} _{-13}$ & 94$^{+6} _{-7}$  & 89$^{+4} _{-5}$   &	                             &        &  0.08$\pm$0.01		\\
    &  0.2417 		&	230$^{+26} _{-25}$ & 98$^{+9} _{-10}$ & 88$^{+14} _{-16}$ &	                             &		&  0.07$\pm$0.02	    \\	
    &  0.2752    	&   203$^{+12} _{-12}$ &                          & 192$^{+17} _{-19}$ & 122$^{+9} _{-9}$ &     	&  0.12$\pm$0.02 		\\
    &  0.2770 		&	186$^{+9} _{-8}$   & 146$^{+12} _{-16}$ &                          & 60$^{+6} _{-6}$  &	-4.6$\pm$0.7  	&  0.126$\pm$0.008 	    \\	
    &  0.2788    	&   67$^{+1} _{-1}$    & 59$^{+2} _{-2}$    & 54$^{+4} _{-4}$    & 37$^{+3} _{-3}$  &        &  0.096$\pm$0.004 		\\
    &  0.2807 		&	63$^{+1} _{-1}$    & 56$^{+2} _{-2}$    & 25$^{+2} _{-2}$    & 26$^{+4} _{-4}$  &  	&  0.097$\pm$0.004      \\	
    &  0.2825    	&   69$^{+2} _{-1}$    & 57$^{+4} _{-4}$    & 31$^{+1} _{-1}$    & 46$^{+6} _{-6}$  &  -4.5$\pm$0.8    	&  0.092$\pm$0.004 	    \\
    &  0.2843 		&	37.3$^{+0.5} _{-0.5}$    & 37$^{+1} _{-1}$    & 27$^{+2} _{-2}$    &                        &  	&  0.073$\pm$0.003 		\\	
    &  0.2861    	&   17.7$^{+0.4} _{-0.4}$    &                          & 12$^{+3} _{-3}$     & 	                    &             	    &  0.058$\pm$0.002 	  	\\
\hline
2020-Jan-27 & & & & & & &\\
    &  0.2734    	&   74$^{+3} _{-3}$    & 73$^{+6} _{-6}$   & 51$^{+3} _{-2}$  & 	&    		&  0.093$\pm$0.006 				   	\\
    &  0.2752    	&   76$^{+3} _{-3}$    & 84$^{+7} _{-6}$   & 51$^{+4} _{-4}$  & 	&    		&  0.094$\pm$0.008 				   	\\
    &  0.2770 		&	142$^{+9} _{-10}$   & 115$^{+9} _{-10}$ & 104$^{+5} _{-5}$ &	&	 -1.8$\pm$0.3  		&  0.17$\pm$0.04  	   			\\	
    &  0.2861    	&   124$^{+5} _{-5}$  & 82$^{+8} _{-8}$   & 66$^{+3} _{-3}$  & 	&    -3.8$\pm$0.3    		&  0.072$\pm$0.003 				   	\\
    &  0.2879 		&	33.6$^{+0.7} _{-0.7}$    & 31$^{+2} _{-2}$   & 28$^{+2} _{-2}$  & 	&	 -1.1$\pm$0.2  		&  0.049$\pm$0.002  	   			\\
\hline
\end{tabular}}
\label{table1}
\end{table*}

\subsection{Pulse scattering}\label{sec3.2}

\begin{figure*}
\centering
\includegraphics[width=150mm]{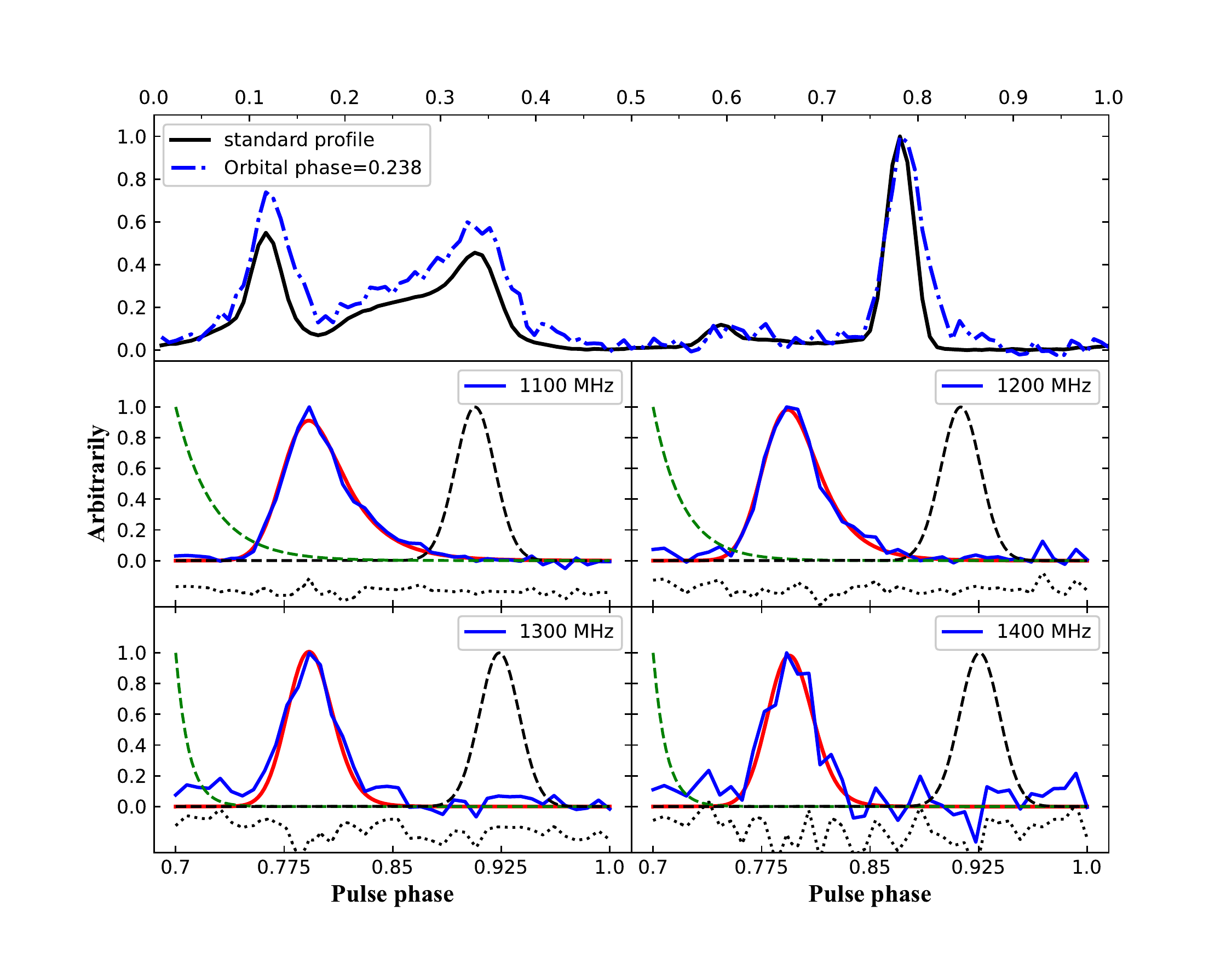}
\caption{Top panel: the averaged pulse profile away from the eclipse region is shown as black solid line and the averaged pulse profile at orbital phase 0.238 is shown as blue dot dashed line. Bottom panels: The main pulse component at orbital phase 0.238 in four subbands (blue lines). Green lines show the best-fit exponential decay function. Red lines show the convolution of the best-fit exponential decay function with a template of the main pulse component (black dashed lines, Gaussian-like function). Black dotted lines show the residuals of fitting.
}
\label{sc_fit}
\end{figure*}

The averaged pulse profile away from the eclipse region is shown in the top panel of Fig.~\ref{sc_fit} as a black solid line. In addition to a narrow main pulse and a broad, asymmetric interpulse~\citep{fruchter90}, we detected a new shallow pulse component at the pulse phase $\sim0.6$. As we move deeper into the eclipse region, the pulse profile becomes significantly wider. As an example, in the top panel of Fig.~\ref{sc_fit}, we show the averaged pulse profile at the orbital phase 0.238 observed on 2019 August 8 as a blue dashed line. In the bottom panels of Fig.~\ref{sc_fit}, we show the main pulse component at the same orbital phase but at four different observing frequencies. Strong evidence of pulse scattering can be observed at low frequencies, especially for the Gaussian-like main pulse component. 

For each 60\,s time window, we split our observing band into four sub-bands and fitted for the pulse scattering time-scale in each sub-band following the procedure described in Section~\ref{sec2}. Taking pulse profiles at the orbital phase 0.238 as an example, in the bottom panels of Fig.~\ref{sc_fit}, we show the best-fit exponential decay function as green dashed lines and the convolution of a Gaussian pulse (dashed black lines) and the exponential decay function as red solid lines. Residuals of the fitting are shown in the bottom as dashed black lines. In Table.~\ref{table1}, we present scattering time-scale measurements as a function of the observing frequency at multiple orbital phases close to the eclipse. We also tried to measure the scattering index by fitting for a power-law, $\tau_{\rm s} = \tau_0\nu ^{\alpha }$, and present the index, $\alpha$ in the second last column of Table.~\ref{table1}. In Table.~\ref{table1}, we only presented scattering time-scales and scattering indices with uncertainty less than 20\% of the measurement, which we consider reliable.

In Table.~\ref{table1} we also present measured DM excesses for these orbital phases. Similar to the flux density variation, we observed large variations of DM close to the egress of eclipse and show different trends in different observations while DMs close to the ingress show significantly less variation. We observed a maximum DM excess of $\sim 0.17$\,pc\,cm$^{-3}$ at the orbital phase 0.277 on 2020 January 27, which corresponds to an electron column density of $\sim5.2\times10^{17}$\,cm$^{-3}$ in the eclipse region ($R_{\rm E}\approx10^{8}$\,m). 
This is consistent with the mean column density of electrons in the outer parts of the eclipsing medium measured by \citet{rt91}.

\section{DISCUSSIONS AND CONCLUSIONS}\label{sec4}

In this paper, we presented highly sensitive observations of PSR~B1957$+$20 centred at 1250\,MHz using FAST. These observations allow us to probe much deeper into the eclipse region of PSR~B1957$+$20 than previous observations. For the first time, we detected pulse scattering and measured scattering time-scales close to the eclipse region. We show that the measured scattering time-scale increases as we go deeper into the eclipse region and reaches a maximum of $\sim0.2$\,ms at $\sim1.1$\,GHz.

The observed pulse broadening can be well described by the convolution of the intrinsic pulse profile and a one-sided exponential decay. We also observed that the scattering time-scale scales with the DM excess roughly as $\tau\propto\Delta{\rm DM^{2}}$. These strongly suggest that scattering is caused by multi-path propagation effects as radio waves traverse a turbulent plasma~\citep[e.g.,][]{Rickett1990}. However, the electron density in the scattering region must be $n_{\rm e} \sim \Delta DM / R_{\rm e}\approx5 \times 10^7$\,cm$^{-3}$, where $R_{\rm E}\approx10^{8}$\,m is the size of the eclipse region. This is comparable with the scattering in the solar corona~\citep[e.g.,][]{ych+12} and at least seven orders of magnitude larger than normal interstellar scattering, so scattering angles and spatial scales will be quite different from normal pulsar observations.

Forward angular scattering due to a medium with variable refractive index is a second moment of the electric field and can be fully characterised by the phase structure function~\citep[see][and reference therein]{cfr+87}. Our measurements of both scattering and electron density provide us the opportunity to investigate the property of eclipsing materials through analysis of the structure function. The angular spectrum of plane waves is $\Phi({\vec{\kappa}})$ where $\vec{\kappa} = k \vec{\theta}$, $k = 2 \pi / \lambda$ is the wave number and $\vec{\theta}$ is the scattering angle. It is the Fourier transform of the auto-correlation function $\rho_e(\vec{s})$. The latter is given exactly by $\rho_e(\vec{s}) = \exp(-0.5 D_\phi (\vec{s}))$. Here $D_\phi (\vec{s}) = \langle (\phi(\vec{r}) - \phi(\vec{r} + \vec{s}))^2\rangle$ is the phase structure function. If the medium is of limited extent, $\phi(\vec{r})$ has an auto-covariance $C_\phi (\vec{s}) = \langle \phi(\vec{r}) \phi(\vec{r} + \vec{s})\rangle$ and it is related to the structure function by
\begin{equation}
\label{eq:sf}
    D_{\phi}(s)=2[C_{\phi}(0)-C_{\phi}(\vec{s})].
\end{equation}
The angular spectrum is approximately Gaussian in shape, so the auto-correlation of the electric field is also approximately Gaussian. Thus we define the width of the auto-correlation as $s_0$ and the width of the spatial spectrum as $\kappa_0 = 1/s_0$, so $\theta_0 = 1/ks_0$ is the width of the angular spectrum. By definition then, $D_\phi(s_0) = 1$.

If the scattering medium is turbulent, the phase structure function will have the form $D_\phi(\vec{s}) \sim s^{5/3}$ in the magneto-hydrodynamic range, where $s_{\rm in} < s < s_{\rm out}$. In a hydrogen plasma, $s_{\rm in} = 6.84 \times 10^5 /n_e^{0.5}$ is the ion inertial scale in m, where $n_{\rm e}$ is in cm$^{-3}$~\citep{ch89}. This gives an inner scale of $s_{\rm in}\approx100$\,m for $n_{\rm e}\approx5 \times$ 10$^7$\,cm$^{-3}$. In general, $s_{\rm out}$ is the extent over which the medium is homogeneous and we expect $s_{\rm out}$ to be of the order of $R_{\rm E}\approx10^8$\,m in our case. For $s < s_{\rm in}$, $D_\phi(s) \sim s^2$ and for $s > s_{\rm out}$, $D_\phi(s) = D_\phi(s_{\rm out})$. 

We can measure $s_0$ using the observed scattering time scale $\tau_{\rm s}$, which is related to $\theta_0$ as $a \theta_0^2 / 2c$. Here $a$ is the distance from the pulsar to the scattering screen ($\sim10^{11}$ cm). For $\tau_{\rm s} =$ 0.2\,ms, we measured $s_0 = 4$ m, which is smaller than the inner scale and smaller by five orders of magnitude than typical ISM values~\citep[e.g.,][]{kcs+13}. At much larger scales, $D_\phi(R_{\rm e})$ can be estimated from the observed $\Delta{\rm DM}$ because $\phi$ and DM are both measures of the electron column density, so $\phi = A\times{\rm DM}$ where, $A=2\pi\nu\cdot[4.15\,{\rm ms}\cdot(\frac{\nu}{1.1\,{\rm GHz}})^{-2}] = 2.3701 \times 10^7$. We assume that the fluctuations in DM are roughly exponentially distributed so Variance(DM) = Mean(DM)$^2$, and $D_{\rm DM}(R_{\rm e}) = 2 \Delta{\rm DM}^2$ = 0.0512.

\begin{figure}
\centering
\includegraphics[width=80mm]{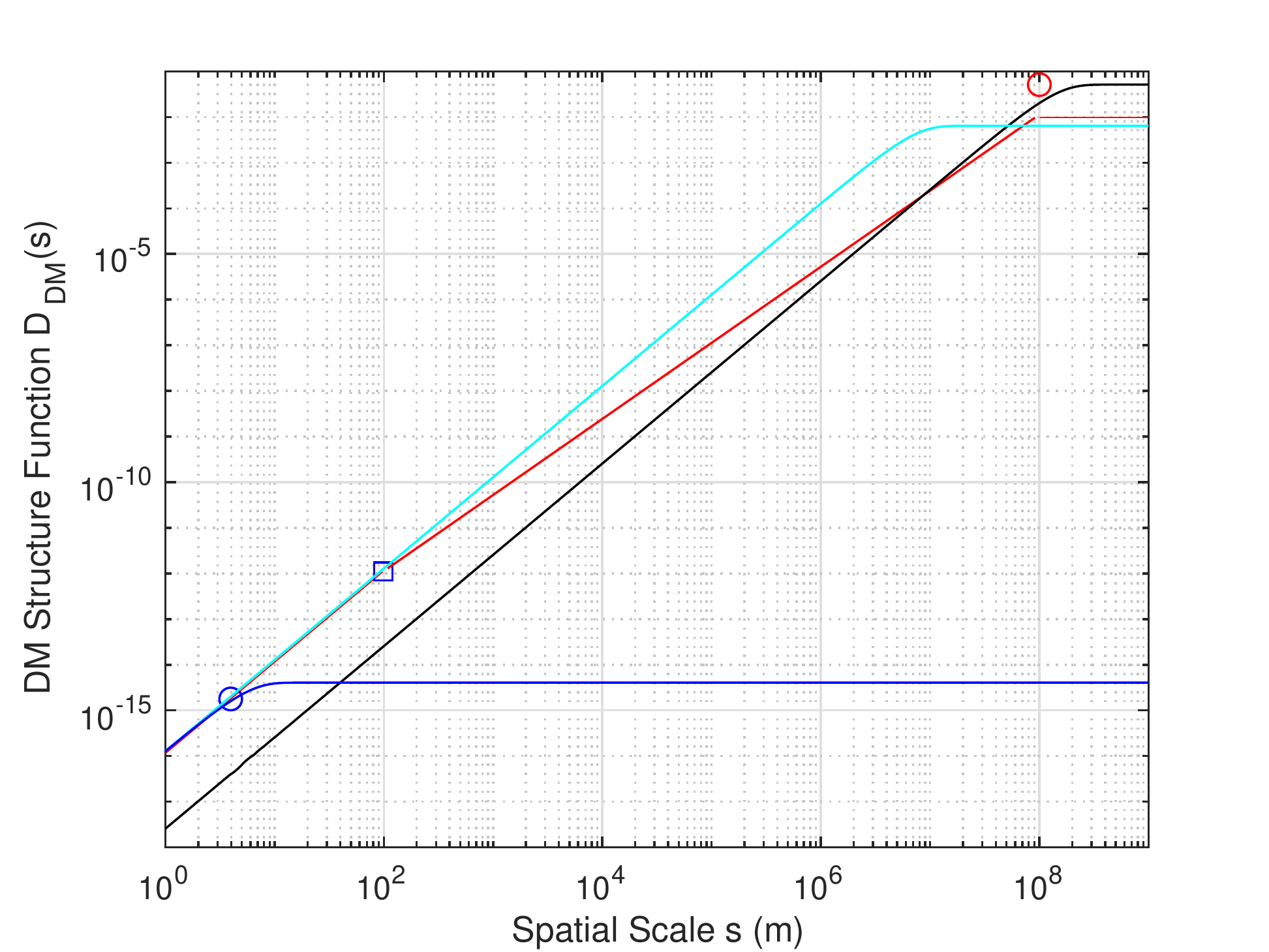}
\caption{Structure function of DM as function of the spatial scale. The red piece-wise linear line represents a Kolmogorov process that matches the observations at $s_0, s_{\rm in}$ and $s_{\rm out}=R_{\rm e}$. The black line represents a process with a Gaussian spatial scale of $R_{\rm e}$ and an RMS electron density $\delta n_{\rm e} = n_{\rm e}$. The cyan line represents a process with a Gaussian scale of $R_{\rm e} /20$ and $\delta n_{\rm e} = n_{\rm e}$. The blue line represents a process with a Gaussian scale of $s_0$ and $\delta n_{\rm e} = 4.3 \times 10^4$\,cm$^{-3}$. The blue circle and square show the measured coherent scale and estimated inner scale, respectively. The red circle shows the estimated $D_{\rm DM}(R_{\rm e})$ based on $\Delta{\rm DM}$.  
}
\label{fig:sf}
\end{figure}

Now we compare $D_{\rm DM}(s)$ of the turbulent plasma model as described above with our observations. The most reliable measurement is $s_0$, so we adjust our model so that $D_{\rm DM}(s_0) = A^{-2}$. The resulting piece-wise linear $D_{\rm DM}(s)$ is plotted as a red line on Fig.~\ref{fig:sf}. We find that our estimate of $D_{\rm DM}(R_{\rm e})$, shown as the red circle on Fig.~\ref{fig:sf}, is a factor of five higher than the model. However, since large variations in both $\tau_{\rm s}$ and $\Delta{\rm DM}$ were measured (Table~\ref{table1}), the assumed density has enough uncertainty to match the model value of $D_{\rm DM}(R_{\rm e})$. For example, it can be brought into agreement if we reduce the assumed root-mean-square (RMS) DM by a factor of $\sim2$.

The electron density fluctuations causing the angular scattering need not to be turbulent. They could have a single spatial scale $s_{\rm ne}$, which can be modeled with a 3D Gaussian auto-covariance for $n_{\rm e}$. The 2D auto-covariance of DM can be obtained by doing the line of sight integrals to obtain DM from $n_{\rm e}$. The 2D structure function can then be obtained from Equation~\ref{eq:sf},
\begin{equation}
    D_{\rm DM}(s) = 2 {\rm Variance(DM)} (1 - \exp(-0.5 (s/s_{\rm ne})^2)).
\end{equation}
When $s_{\rm ne} = R_{\rm e}$, the line of sight integration is degenerate and Variance(DM) = $\delta n_{\rm e}^2 R_{\rm e}^2$. When $s_{\rm ne} \ll R_{\rm e}$, the line of sight integration gives Variance(DM) = $\sqrt{2 \pi}\,\delta n_{\rm e}^2 R_{\rm e} s_{\rm ne}$.

First we model the large scale fluctuations in DM as Gaussian, setting $s_{\rm ne} = R_{\rm e}$ and $\delta n_{\rm e} = n_{\rm e}$. This $D_{\rm DM}(s)$ is plotted on Fig.~\ref{fig:sf} as a black line. One can see that it fails to provide enough scattering as the predicted structure function is lower by nearly two orders of magnitude at $s_0$. In order to explain the observed angular scattering with this model, one needs to reduce $s_{\rm ne}$ by a factor of $\sim20$, which is shown with a cyan line in Fig.~\ref{fig:sf}. Although technically possible, we should have observed such rapid and large variations in $\Delta DM$ close to the eclipse, so it seems unlikely.

\citet{thompson94} have suggested that strong scattering can be caused by Langmuir turbulence with the electron density fluctuations at extremely short scales. 
They noted that such short-scale plasma fluctuation need to be continuously regenerated and a promising source of energy is the relativistic particles from the pulsar wind. 
We have modeled this suggestion as a Gaussian process with a scale of $s_0$. It would require an RMS electron density of $\delta n_{\rm e}\sim4.3\times10^{4}$\,cm$^{-3}$ to create the observed angular scattering. This structure function is shown as a blue line on Fig.~\ref{fig:sf}, which fails to explain large scale fluctuations in DM. If the scattering medium had two static Gaussian components, one of scale $R_{\rm e}$ and another of scale $s_0$, we would not be able to distinguish that from a single turbulent process. On the balance, we prefer the turbulent model as it fits the observations with minimal assumptions.

With measurements of the scattering time-scale at multiple frequencies, we tried to fit for a power-law, $\tau _{\rm s} = \tau_0\nu ^{\alpha }$, and measure the scattering index $\alpha$. Because of the steep spectrum of PSR B1957+20, scattering time-scale measurements at high frequency subbands are poor with large uncertainties. Therefore, $\alpha$ is not well constrained. However, for those with relatively small uncertainties, we find that $\alpha$ varies from $\sim-1.5$ to $-4.6$. Although an exponent of $-4.4$ is predicted for a scattering screen with Kolmogorov turbulence~\citep[e.g.,][]{LG76,rickett77}, it would not be expected in our observations because the structure function is quadratic in the vicinity of $s_0$. Thus the expected exponent is $-4.0$ regardless of whether it comes from a Gaussian process or a turbulent process which has become heavily damped by the scale of $s_0$. 

Our highly sensitive observations enable us to probe deeper into the eclipse region. The total eclipse duration is measured to be $0.029\pm0.003$ ($16\pm2$ minutes) and $0.041\pm0.004$ ($22\pm3$ minutes) at 1250\,MHz on 2019-Aug-08 and 2020-Jan-22, respectively.   
This is significantly shorter than the previous Arecibo measurement of 0.06$\pm$0.01 ($33\pm8$ minutes) at 1400\,MHz ~\citep{rt91}.  
Both DM and flux density at the egress are highly variable, and we observed a sudden rise of flux density at the orbital phase of 0.263, deep into the eclipse region. This, again, shows that the trailing tail of eclipsing materials is highly clumpy~\citep{TB91}. While the egress duration is uncertain due to large flux density variations, we show that the ingress duration is much shorter than previously measured. Combining our measurements with previous low frequency measurements, we find that a single power-law cannot fit the frequency dependence of eclipse duration and the duration at 1.25\,GHz is significantly shorter than previously expected.

Currently, the eclipsing mechanism and properties of eclipsing materials are still not well understood. \citet{Kudale20} discussed several eclipse mechanisms for the case of PSR~J1227-4853 and concluded that the eclipse was likely to be caused by cyclotron-synchrotron absorption. \citet{kansabanik21} modeled the broadband radio spectrum of PSR~J1544+4937 at its full eclipse phase and showed that synchrotron absorption by relativistic electrons is favoured. In the case of PSR~B1957+20, based on early observations \citet{thompson94} suggested that a favoured model is cyclotron or synchrotron absorption by plasma embedded in the pulsar wind combined with pulse smearing at high frequency (i.e., 1.4 GHz). More recently, low frequency observations carried out by \citet{Polzin20} supported absorption or nonlinear scattering eclipse mechanisms. Our observations provided the first evidence of strong scattering close to the eclipse of PSR~B1957+20 and confirmed that strong scattering is playing an important role at around 1.4\,GHz. However, we show that even after considering the scattering the averaged flux density decreases as we go deeper into the eclipse, which suggests that other eclipsing mechanisms are also involved. Simultaneous measurements of pulsed and continuum flux densities of PSR~B1957+20 close to the eclipse at around 1.4\,GHz will help us better understand the contribution of pulse scattering.

Neutron stars in binary systems or other extreme environment have been suggested to be possible origins of fast radio bursts~\citep[FRBs, e.g.,][]{zha17,mm18,lbg20}. In these models radio emission from neutron stars are expected to propagate through their surrounding dense and ionised medium, which is similar to the case of eclipsing binaries. Observationally, strong pulse scattering have been observed in both non-repeating~\citep[e.g.,][]{osr+19,qsf+20} and repeating FRBs~\citep{nal+21}. There are also increasing evidence of dense ionised medium surrounding FRB sources, especially those active repeating ones~\citep[e.g.,][]{msh+18,fly+22,acb+22,dfy+22}. Studies of the scattering of FRB~190520B implied distance between the FRB source and dominant scattering material is $\lesssim100$\,pc~\citep{occ+22}. Our detection of strong scattering close to the eclipse of PSR~B1957+20 show that ionised medium surrounding neutron stars in eclipsing binary systems can be highly turbulent and responsible for the observed scattering. Future observation of pulse scattering of FRBs may provide additional information about their origin.

\section*{Acknowledgments}
We thank Stefan Os{\l}owski and Congyao Zhang for useful discussions. This work made use of the data from FAST (Five-hundred-meter Aperture Spherical radio Telescope).  FAST is a Chinese national mega-science facility, operated by National Astronomical Observatories, Chinese Academy of Sciences. SD is the recipient of an Australian Research Council Discovery Early Career Award (DE210101738) funded by the Australian Government. This work was funded by the National Natural Science Foundation of China (Nos.U1731238, 11565010), Foundation of Science and Technology of Guizhou Province Nos. (2016)4008, (2017)5726-37 and Foundation of Guizhou Provincial Education Department (No. KY(2020)003). This work is supported by the National Natural Science Foundation of China under Grand No. 11703047, 11773041, U2031119, and 12173052. ZP is supported by the CAS “Light of West China” Program. Shijun Dang is supported by Guizhou Provincial Science and Technology Foundation (Nos. ZK[2022]304)

\section*{Data Availability}
The observations from the FAST radio telescope are publicly available from \url{https://fast.bao.ac.cn/} after an 12 month embargo period. Note that all original data used in this paper are out of this embargo period and are available upon request. Processed data products used in this paper are available from \url{https://www.scidb.cn/en/doi/10.11922/sciencedb.01629}. 


\bibliographystyle{mnras}
\bibliography{references} 







\bsp	
\label{lastpage}
\end{document}